\documentstyle[prl,aps,preprint,tighten]{revtex}
\newcommand{\XOR}{\oplus}
\newcommand{\OR}{\vee}
\newcommand{\AND}{\wedge}

\draft
\title{Algorithmic Mapping from
Criticality to Self Organized Criticality}
\author{F. Bagnoli\cite{fb}}
\address{Dipartimento di Matematica Applicata, 
Universit\`a di Firenze, via S. Marta,
3 I-50139, Firenze, Italy.}
\author{P.Palmerini}
\address{Dipartimento di Fisica, Universit\`a di Firenze, Largo E. Fermi, 2, I-50125
Firenze, Italy.}
\author{R. Rechtman\cite{rr}}
\address{Centro de Investigaci\'on en Energ\'{\i}a, 
UNAM, Apdo. Postal 34, 62850 Temixco Mor., Mexico.
}
\date{\today}
\begin{document}

\maketitle

\begin{abstract}                
Probabilistic cellular automata are prototypes of non equilibrium critical
phenomena. This class of models includes among others
the directed percolation problem (Domany Kinzel
model) and the dynamical Ising model. The critical properties of these
models are usually obtained by fine-tuning one or 
more control parameters, as for
instance the temperature. 
We present a method for the parallel evolution of the
model for all the values of the control parameter, although its implementation
is in general limited to a fixed number of values. This algorithm facilitates
the sketching of phase diagrams and can be useful in deriving the critical
properties of the model.
Since the criticality here emerges from the asymptotic distribution of some
quantities, without tuning of parameters, our method is 
a mapping from a probabilistic cellular automaton with
critical behavior to a self organized critical model with the same critical
properties.
\end{abstract}



\section{Introduction}
Recently, several papers \cite{Sornette1,Maslov,PMB,Sornette2,Grassberger}
have appeared discussing the relations between
 self organized criticality (SOC) \cite{SOC}  and 
usual critical phenomena. Some of them \cite{Grassberger} stress the fact that 
one can reformulate classic critical systems (namely, directed percolation)
in a way indistinguishable from SOC, while others \cite{Sornette1} focus
on the role of the control and order parameters.  

We started our investigation from the  observation~\cite
{Grassbergerorig} 
that one can express the problem of directed 
site and bond
percolation~\cite{DP} in a form reminiscent of the invasion percolation
process~\cite{Invasion} or the Bak-Sneppen self organized model~\cite
{BakSneppen}. The advantage of this formulation is that the critical value
of the percolation probability does not need to be adjusted carefully, but
instead emerges from the probability distribution of a set of continuous
variables, while the original model is defined in terms of Boolean
variables.

The directed percolation problem can be formulated in terms of probabilistic
cellular automata (PCA)~\cite{DK}. PCA are very general models that
include for instance the kinetic Ising model \cite{Georges}. 

In this paper we show how any critical PCA may be mapped into
a SOC model. The mapping is presented constructively 
in Sec.~\ref{section:fragments}. 
It can also be considered
as a multi-site coding technique~\cite{multispin}, particularly adapted to 
probabilistic systems (where usual multi-site performs badly). 
From a computational point of view, this algorithm allows a quick determination of
phase diagrams and computation of critical properties. 
In Sec.~\ref{section:examples} we apply this method to the 
study of the Domany-Kinzel model of directed percolation
and to the two dimensional Ising model. 

On the other hand, the mapping implies that to each PCA corresponds
a SOC model defined in a high or infinite dimensional space. This
correspondence can give some insight in the nature of the SOC
phase, as addressed in Sec.~\ref{section:SOC}. 
We end with some conclusions and perspectives.

\section{The fragment method}
\label{section:fragments}
We deal with probabilistic cellular automata, i.e., discrete models 
defined on a lattice. 
Let us consider explicitly the one dimensional Boolean
case. A configuration  at time $t+1$ is obtained from the 
configuration at time $t$ by applying in parallel a probabilistic rule
to each site. 
The rule is implemented on a computer by comparing (pseudo) 
random numbers with a
certain number of fixed parameters (probabilities). One can think of PCA as
the evolution of a deterministic discrete system on a random quenched field
(the set of random numbers). 

For
simplicity, we refer to the directed site percolation problem in 1+1
dimensions, where the higher $p$, the higher the probability of percolating.
In this case one can visualize the random field as the height of a corrugated
landscape, and $p$ as the water level. There will be percolation if the
water is able to percolate on the corrugated landscape, i.e., if the plane at
height $p$ is not completely blocked (in this directed model water is
forbidden to back-percolate). For each value of $p$, we denote with a one the 
sites that are wet, and with a zero those that are dry.

One can stack a set of planes, and let them evolve in parallel. 
We can read the state of a certain site  
for all values of $p$ as a vector
of ones and zeroes, each component being labeled by $p$. 

In the initial configuration sites are either wet or dry, independently
of $p$. Thus, all vectors are either filled with ones or with zeroes.  
Going on with
the percolation process, a component $p$ of the vector at 
a certain site and time $t+1$
will be wet if there is at
least a wet component at the same height among its neighbors at time $t$, 
and if the height of the
random field at that position is less than $p$. One can easily express this in
computer language. Each component corresponds to a bit in a computer word. With
words of $n$ bits, $p$ can assume the values $0/n, 1/n, \dots, i/n, \dots
(n-1)/n$. 
The (bitwise) OR of the words in the neighborhood gives a one for 
all values of $p$ for which there is at least one wet neighbor. 
Given a random number $r$ in that site, all
planes with $p>r$ have the possibility of percolating. This is expressed in
computer terms by taking a word $R(r)$ filled with zeroes up to a fraction
$r$ of bits  and  then with
ones, 
and performing the AND of $R(r)$ with the previous word.
Iterating this procedure, we get in the last line of the lattice (say at
time $T$) a set of partially filled words. If at time $T$ a word has the bit
number $k$ equal to one, this means that for $p=k/n$, water would have
percolated to that site (given the set of random numbers).

The procedure can be generalized to words of arbitrary length. In the limit
$n \rightarrow \infty$, the Boolean vectors become the characteristic 
functions 
of subsets of the unit interval, that we call fragments. 
The manipulation of fragments  is not limited to this bitwise
implementation, as we shall see in the following.

The fragment expressions do not depend explicitly on the control parameter 
$p$. The critical value of $p$ and the critical scaling law of the order
parameter are obtained {\em a posteriori\/}, from the distribution of fragments.

Let us now formalize these concepts.
For simplicity we refer to  the Domany-Kinzel (DK) model~\cite{DK}, which 
is a simple one dimensional PCA.
We denote with $x_i^t=0,1$ the state of a
site $i$ at time $t$, $i=0, \dots, L-1$, with $L$ the size of the lattice.
We shall simplify $x^{\prime}=x_i^{t+1}$, $x_\pm =
x_{i\pm 1}^t$. All space index operations are modulo $L$ (periodic boundary
conditions). 
The evolution rule may be written as 
\begin{equation}
x^{\prime }=[r<p](x_{-}\oplus x_{+})\vee [r<q]x_{-}x_{+}  \label{DK}
\end{equation}
where $\oplus$ represents the eXclusive OR operation (sum modulus two), $%
\vee$ the OR operation, and the multiplication (or $\AND$) stands for the AND
operation, with the usual priority rules. The control parameters $p$ and $q$
are fixed, and $r=r_i^t$ is a random number uniformly distributed between
zero and one, and where $[\mbox{\it logical expression}]$ is one if {\it logical
expression} is true and zero otherwise~\cite{Iverson,Knuth}.

In the case of directed site percolation $p=q$, Eq.\ (\ref{DK}) can be
rewritten as  
\begin{equation}
x^{\prime}(p)=[r<p]  (x_-(p) \vee x_+(p)).  \label{site}
\end{equation}
where we emphasize the dependence of $x$ on $p$.

The fragment approach consists in reading $x_i^t(p)$ as the 
value of the characteristic function of the fragment $X_i^t$ at $p$. 
The expression $[r<p]$ is the characteristic function of a fragment $R(r)=[r,1)$.

Equation~(\ref{site}) in terms of fragments is 
\begin{equation}
X^{\prime}=R(r)(X_- \vee X_+)   \label{SITE}
\end{equation}
that does not depend on $p$. The Boolean functions AND, OR, XOR, NOT
correspond to the set operations intersection, union, symmetric difference and
complement, respectively.
We shall use the same symbol for Boolean and 
set operations. 
The initial configuration is independent of $p$, this means that $X_i^0$ are
either the empty set or the unit interval, according with $x_i^0$. 
Applying the set operations we obtain the asymptotic  fragments $X_i^\infty$.

For a given value of $p$, $x_i^t(p)$ is one if the point $p$ belongs to the
fragment $X_i^t$ and zero otherwise. Thus we can obtain the asymptotic value of
$x_i^t(p)$ from the asymptotic fragments $X_i^t$, that evolved without 
an explicit dependence on $p$.  If some
function of the $x_i^t$ exhibits a phase transition in
correspondence of a critical value $p_c$, this behavior can be extracted
from the asymptotic fragments. For instance,  
the density $\rho$
\begin{equation}
\rho(p) =  \frac{1}{L} \sum_{i=1}^{L} x^T_i(p)
\end{equation}
is proportional to the number of fragments the point $p$ belongs to,
\begin{equation}
\rho(p) =  \frac{1}{L} \sum_{i=1}^{L} [p \in X^T_i].
\end{equation}

The above procedure can be applied to all probabilistic 
cellular automata.
The practical recipe for the implementation is
\begin {description}
\item{\bf PCA:} Express the model  as a probabilistic cellular automaton
whose evolution rule only uses Boolean expressions, and
convert the control parameters ($p_1,p_2,\dots, p_m$) to expressions like $%
[r_k<p_k]$ where $p_k$ appear alone on the right side. 
\item{\bf Fragments:} Replace the variables $x_i^t$ with fragments 
$X_i^t\subseteq [0,1)^m$,
and substitute $[r_k < p_k]$ with $R(r_k)$ ($%
[r_k>p_k]$ with its complement $\overline R(r_k)$). 
The initial configuration $x_i^0$ is replaced by $X_i^0 = \overline R(x_i^0)$.
\item{\bf Implementation:} Implement the fragments as arrays of bits (the simplest approach) or
as sparse vectors (see later) and iterate the rule.
\item{\bf Criticality:} The asymptotic distribution of 
fragments gives the critical properties
(control parameters and exponents) of the original model. 
\end {description}

Let us illustrate separately each of the previous points. 

\subsection{PCA}
The evolution rule is generally expressed by means of transition
probabilities. Note, however, that the transition probabilities do not 
completely characterize
 the problem for the damage spreading
transitions~\cite{Bagnoli:damage}, since there are many ways of actually
implementing the probabilistic choices in a computer code. The general approach
 for deriving a Boolean expression from transition probabilities 
is to write formally the future value of the
dynamical variable (the spin) as a function of the spins in the neighborhood and
of the transition probabilities converted to random Boolean variables, i.e., 
 $x' = f(x_-, x_+, \dots, [r<p_1], [r<p_2], \dots)$.
Then there are several
ways~\cite{ComplexityOfBoolean,Bagnoli:Boolean} of
expressing a Boolean function using a set of standard Boolean operations like
AND, OR, XOR, NOT. Clearly, one should spend some effort in reducing the length
of   the resulting expression.
Sometimes (see the Ising
model in Sec.~\ref{section:examples}) 
one has to transform from $[r<f(p)]$ to something
like $[f^{-1}(r)<p]$ (or more complex expressions).

\subsection{Fragments} 
The method can be applied to any number of parameters. In the case of 
$m$ parameters $p_1,\dots, p_m$ the fragments are subsets of the 
$m$-dimensional unit hypercube.  
 For instance, in the general DK model, Eq.\ (\ref{DK}),
there are two control parameters ($p$ and $q$) and the fragments are
subset of the unit square.  In this way it is possible to draw a sketch
of a phase diagram, in just one simulation.

However, if one is interested in crossing the critical surface along one
line (see the computation of critical exponents in 
Sec.~\ref{section:criticality}), 
 one has to express the parameters as functions of a single
variable, say $s$, and transform the expressions accordingly. For
instance, the directed bond percolation problem corresponds to the curve $%
q=p(2-p)$, which can be expressed as $p=s$ and $q=s(2-s)$. The corresponding
expression is
\begin{equation}
x^{\prime }=[r<s](x_{-}\oplus x_{+})\vee [1-\sqrt{1-r}<s](x_{-}x_{+})
\label{bond}
\end{equation}
which gives the following fragment expression
\begin{equation}
X^{\prime }=R(r)(X_{-}\oplus X_{+})\vee R(1-\sqrt{1-r})(X_{-}X_{+}).
\end{equation}

\subsection{Implementation}
 \label{item:layer} 
The simplest way of implementing a fragment on a computer is 
by means of an array of $n$ bits and using bitwise Boolean operations. Generally
one uses computer words (32 or 64 bits) 
for efficiency, but it is possible to use 
multiple words to increase the sampling frequency of probability.  
 The numerical advantage over other multi-spin approaches \cite{multispin}
is the use of just one random number for all the $n$ simulations. 
Referring to a $p$-layer as a cut of the fragment space-time configuration with 
a given value of $p$, we see that 
the layers at different
$p$ are not independent, since they use the same random numbers.
 The influence of 
these correlations is discussed in Sec.~\ref{section:correlations}. 
 
One can increase the sampling frequency around the region of interest 
(for instance the critical region) by 
appropriately
defining the correspondence of the bits with the values of $p$. This affects
the way $R(r)$ is implemented.
 With the fragment method
it is still
possible to perform simulations starting from a single site (Grassberger
method) keeping track of nonzero fragments. The method is powerful
if one uses a small interval around the critical point, so that all
clusters for various $p$ are similar. 
When using two parameters (say $p$ and $q$) one has to implement differently
expressions like $[r<p]$  (fill the unit square in the $p$ direction from $r$
to 1) from $[r<q]$ (fill the unit square in the $q$ direction from $r$
to 1).   

The alternative approach in representing fragments 
consists in keeping track of the starting and ending
points of all segments that form a one dimensional fragment. The approach is
very similar to the treatment of sparse matrices, so we call it the sparse
fragment method. The rules of combining sparse fragments are more complex than
above. On the
other hand in this way one has infinite precision, 
which comes useful in finding the critical behavior as explained
in the Sec.~\ref{section:correlations}.

In general a fragment is formed by  just one segment if the evolution 
rule can be expressed
using only AND and OR, while for instance the XOR between two
overlapping fragments causes holes. 
As an example, the site percolation rule Eq.\ (\ref{site}) can be implemented
as sparse fragments by considering the evolution of the lower extremum 
$a = a_i^t$ of the segments $[a,1]$ as~\cite{Roux,Sornette1,Grassberger}
\begin{equation}
a' = \max(r , \min(a_-, a_+)).
\label{SOCsite}
\end{equation}

Sometimes the problem can be reformulated without XOR operations. 
For instance, the bond percolation problem~(\ref{bond}) 
can be rewritten as~\cite{Grassberger,Roux,Halpin} 
\begin{equation}
x^{\prime }=([r_{-}<p] x_{-})\vee ([r_{+}<p] x_{+})
\label{Grassberger:bond}
\end{equation}
with two random numbers per site. The evolution of this rule can be easily
implemented using sparse fragments. 

\subsection{Criticality}
\label{section:correlations}
\label{section:criticality}
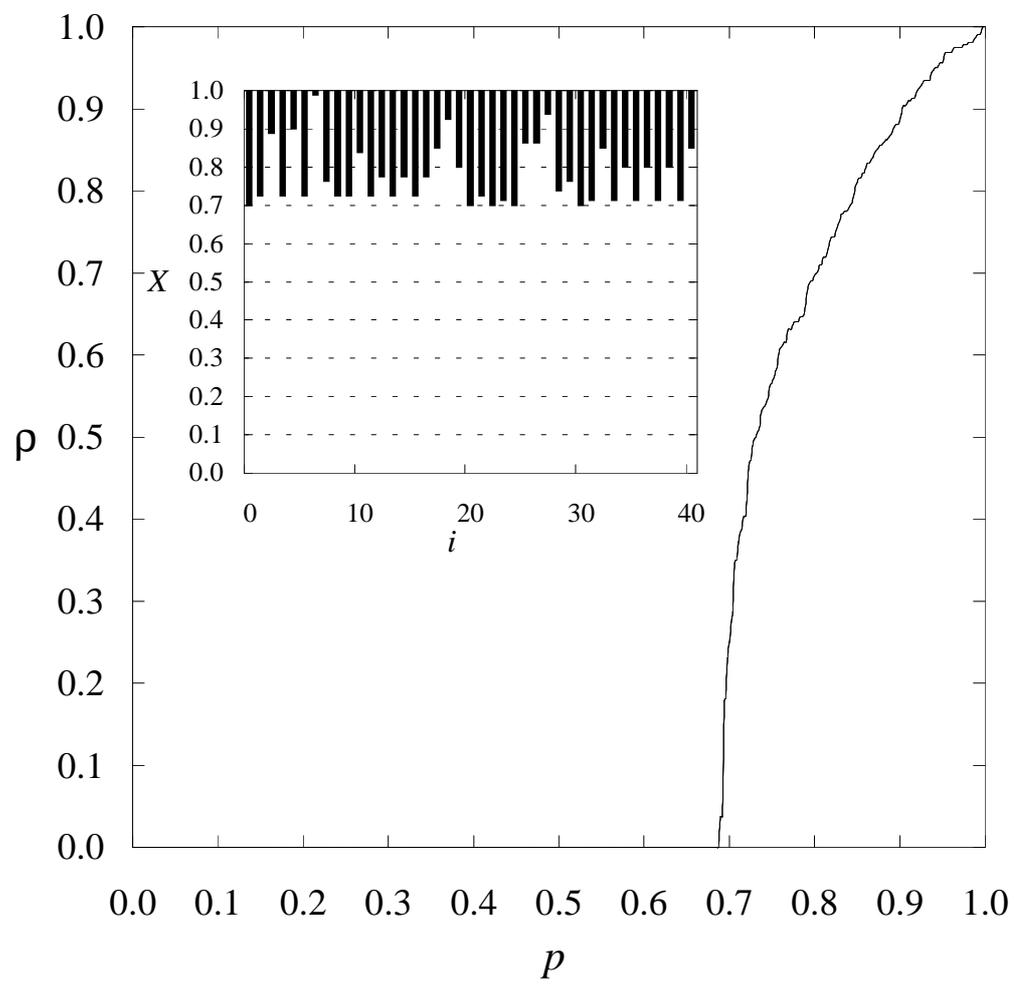
\begin{figure}
\caption{\label{fig:or}The density $\rho $ vs. the control parameter $p$ for
the directed site percolation problem, Eq.\ (\protect\ref{site}). The
inset shows asnapshot of the first 40 segments ${\bf X}^t$,
Eq.\ (\protect\ref{SITE}) after $t$ time steps.. One
simulation with $L=320$ and $t=1000$. The resolution is 480 bits.}
\end{figure}

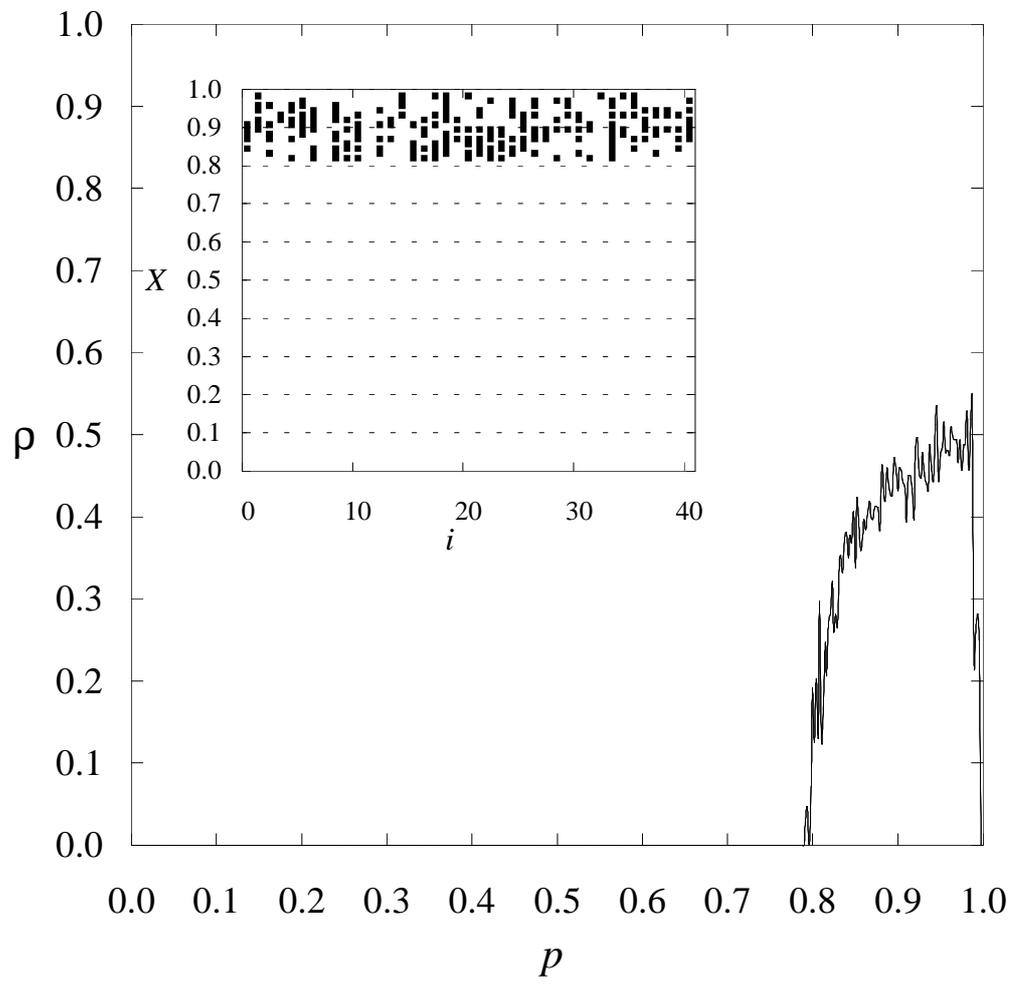
\begin{figure}
\caption{\label{fig:xor}The density $\rho $ vs. the control parameter $p$ for
the XOR dilution, Eq.\ (\protect\ref{DK}) with $q=0$. The inset shows
a snapshot of the first 40 segments ${\bf X}^t$ after $t$ time steps..
One simulation with $L=320$ and $t=1000$. The resolution is 480 bits.
Notice that the simulation
reproduces well also the point $p=1$, for which $\rho =0$.}
\end{figure}

The critical properties of the original model are obtained from the 
asymptotic distribution of fragments.
The fragment method introduces strong
correlations among $p$-layers as also noted in Ref.~\cite{Grassberger}. 
One can exploit these correlations considering differences 
in the $p$ direction. If the patterns for different $p$-layers have similar
sizes, the fluctuations cancel out. This happens in general if the rule 
does not contain XOR (see Fig.\ \ref{fig:or}). On the other hand, the XOR
generally implies strong variations of clusters with $p$, so that the
fluctuations can in principle be wider than uncorrelated simulations
(see for instance Fig.\ \ref{fig:xor}). 

 A powerful method for the computation of critical quantities 
 exploits the scaling relation 
\begin{equation}
m(p,t)=\alpha ^{-\beta /\nu }m(\alpha ^{1/\nu }(p-p_c)+p_c,\alpha t)
\end{equation}
numerically solving it for the unknown $\beta $, $\nu $ and $p_c$.
This is an easy task for the sparse fragments approach, 
since one can obtain $m(p,t)$
and $m(p^{\prime },\alpha t)$ with $p^{\prime }=\alpha ^{1/\nu }(p-p_c)+p_c$
for each value of $\nu $ and $p_c$. For the bit approach one has to
compute the value of the exponents and $p_c$ 
so to make all data collapse on a single
(smooth) curve. This can be performed near $p_c$ 
approximating the curves with polynomials (or any other fitting function),
and minimizing the $\chi ^2$ of the regression.

\section{Applications}
\label{section:examples} 
In this section we show some results of the
fragment method applied to classical problems: the determination of the
phase diagram and critical properties of the Domany-Kinzel model, and the
two-dimensional Ising model.

The first example is the one-dimensional directed site percolation, i.e., the
line $q=p$ of the DK model, Eq.\ (\ref{site}). 
A snapshot of  part of the asymptotic fragment
configuration is shown in Fig.\ \ref{fig:or},  with the plot
of the density $\rho(p)$. As illustrated above, 
if one computes the XOR dilution (the line $q=0$ of the DK
model, Eq.\ (\ref{DK})), the fragments decompose into several segments, 
as shown in the inset
of Fig.\ \ref{fig:xor}. Correspondingly, the integrated density converges 
slowly to a smooth curve.

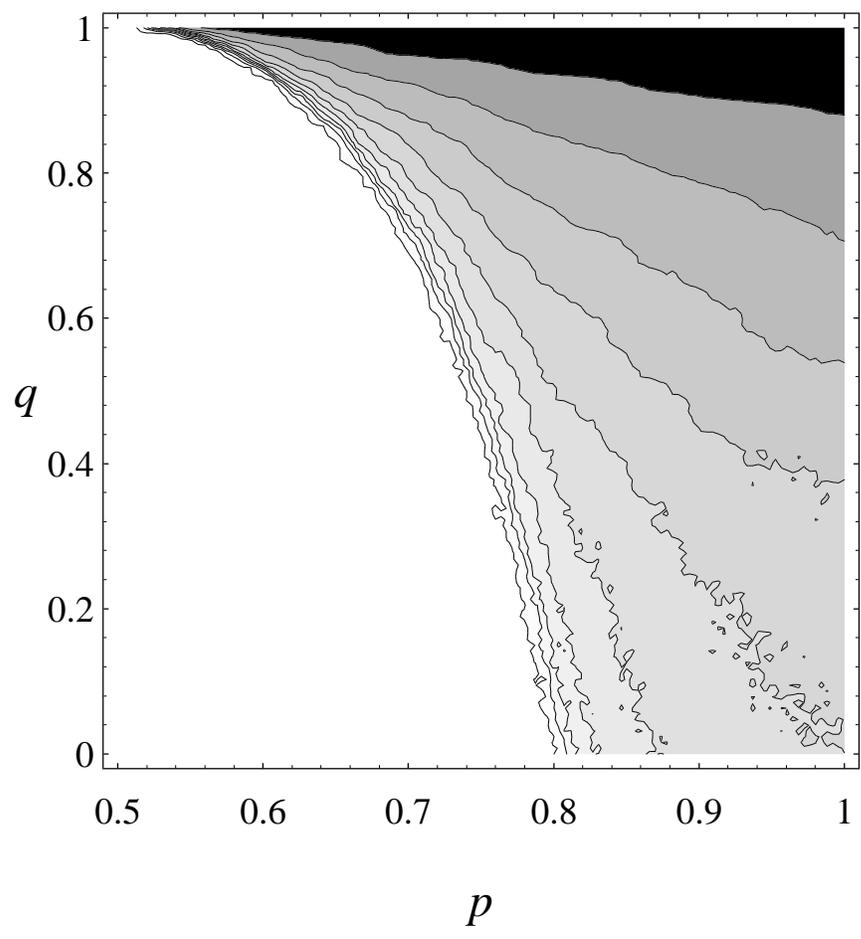
\begin{figure}
\caption{\label{fig:DK}The contour plot of $\rho (p,q)$ of the Domany-Kinzel
Model, Eq.\ (\protect{\ref{DK}}), for a lattice of $L=2000$ sites and $t=4000$.
The resolution is $128\times 128$ bits. White corresponds to $\rho=0$ and
the contour lines are drawn at 0.1 intervals.}
\end{figure}

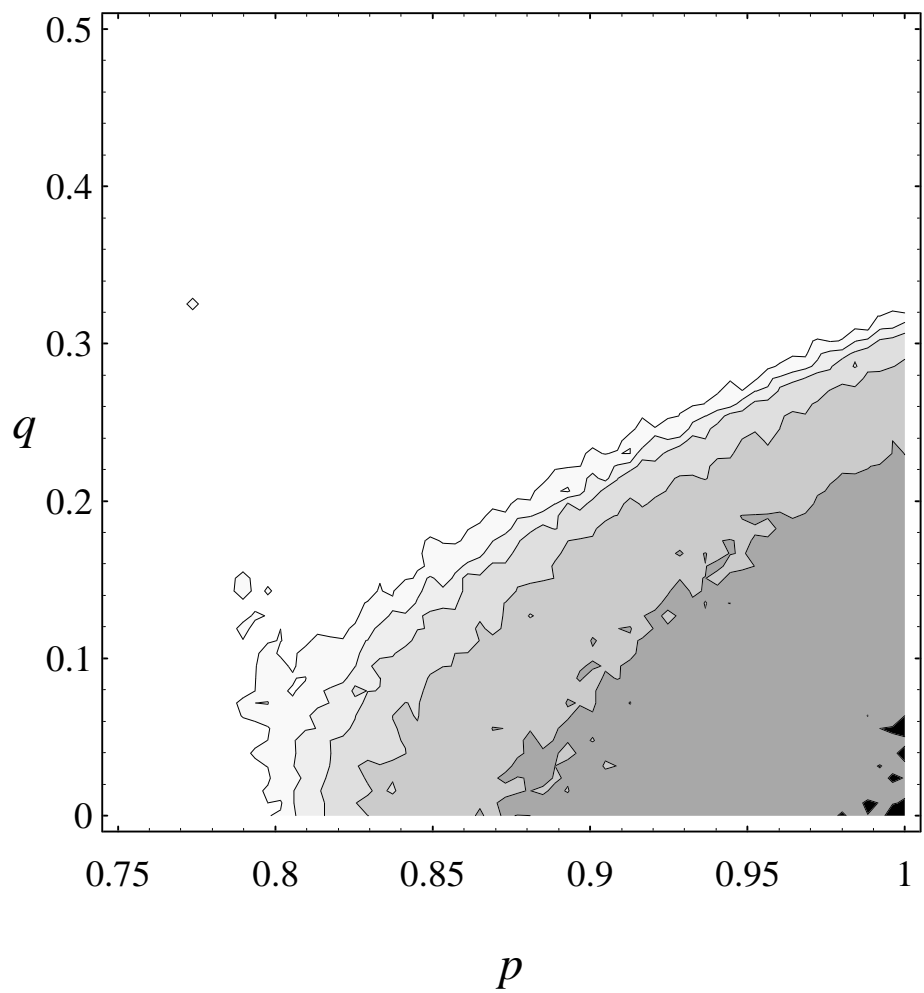
\begin{figure}
\caption{\label{fig:DKH}The contour plot of the asymptotic value of the
distance between two replicas of the system evolving under the same
realization of the noise. The parameters are those of the previous figure.}
\end{figure}

The complete phase diagram of the DK model can be obtained in just one
simulation by iterating two-dimensional fragments. 
The plot of the asymptotic density $\rho(p,q)$ is shown
in Fig.\ \ref{fig:DK}. It
compares well with those obtained with other 
methods~\cite{DK,Bagnoli:damage,Bagnoli:PRG}.
From the convergent behavior
of the contour lines, the position $(p=1/2, q=1)$
of the discontinuous
transition for the density is clearly indicated. 
Near the corner $(p=1, q=0)$ the surface becomes irregular:
this is due to the prevalence of the XOR in eq.~(\ref{DK}).  
One can also investigate the chaotic phase of the DK 
model by iterating two fragment configurations 
with the same random numbers. The Hamming distance between two 
replicas with a given value of $p$ and $q$ is the $(p,q)$-component
of the density of the XOR between the two asymptotic fragment
configuration. A plot of the resulting Hamming distance is shown 
in  Fig.\ \ref{fig:DKH}. Here one can notice a trace of the
density phase boundary (near $(p=0.8, q=0)$), due to the
critical slowing down. 

For the directed site percolation problem we found
$p_c=0.7055(4), \beta =0.21065(5), \nu =1.7195(5)$
for a system of size $10^6$, in the interval $0.7<p<0.71$ for different values
of $\alpha=1024, 2048, 4096, 8192$. 
The agreement with previous measurements~\cite{DP} is satisfactory. 
Moreover, we want to stress that these 
values were obtained with data coming from simulations of less than 20 minutes 
of CPU time on a 150 MHz PC running Linux~\cite{linux}.  

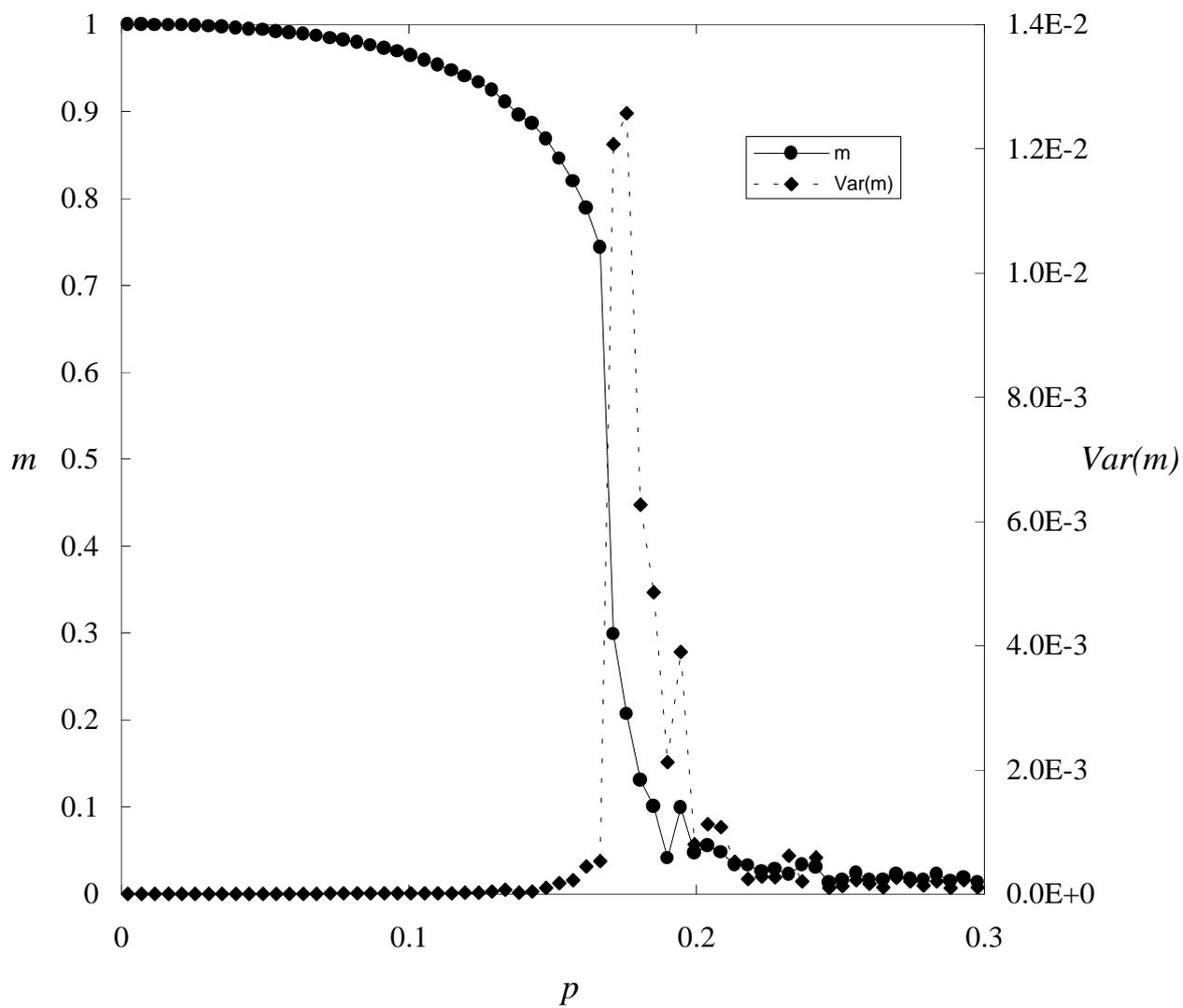
\begin{figure}
\caption{\label{fig:Ising}The magnetization $m$ and the susceptibility
$Var(m)$ (fluctuation of the magnetization) for a 2-dimensional Ising model
of size $L=100\times 100$, averaging over 10 samples every 1000 time steps
after a transient of 10000 time steps. The resolution is 32 bits. }
\end{figure}

As a second application, 
we consider now the kinetic version of an equilibrium system,  the
two dimensional Ising model with heat bath dynamics \cite{Herrmann,Wang}.
In Appendix A we show how to express the evolution equation of this 
model as a totalistic PCA, and how to translate its evolution 
in fragment language. 
In Fig.~\ref{fig:Ising} we show the plot of the magnetization $m(p)$
with respect to $p=\exp(-2J)$ for an
 Ising  model with reduced interaction constant $J$.
The transition is well characterized by plotting the 
second moment (standard deviation)
of the magnetization as a function of $p$. We found $p_c = .172 \pm 0.002$
and $\beta = .11 \pm .002$, in good agreement with the exact values 
$p_c = (\sqrt{2} -1)^2$ and $\beta = 1/8$.

\section{Criticality and Self-Organized Criticality}
\label{section:SOC}
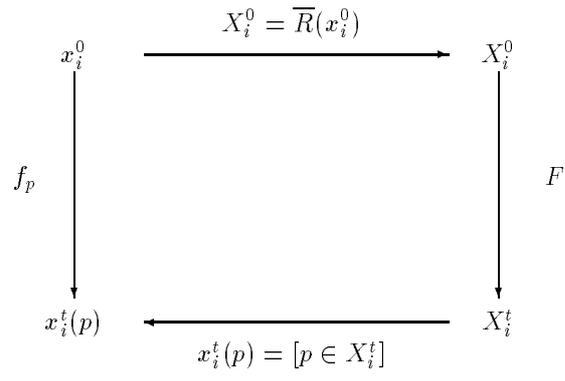
\begin{figure}
\caption{\label{figure:diagram}The diagram showing the mapping from 
criticality to self organized criticality.}
\end{figure}

We have shown how the DK model and the Ising model, 
which are PCA with critical behavior, may be mapped into fragment models 
with no control parameter, that is models that show self organized 
criticality (SOC). It is evident that the fragment method may be applied
to any critical PCA. This result is summarized in the diagram of 
Fig.\ \ref{figure:diagram}. The state $x_i^t(p)$ may be obtained
by evolving the PCA with a given $p$ (labeled by $f_p$ in the diagram) 
or by building the fragments
$X_i^0$ and evolving them with the fragment method ($F$ in the diagram)
that does not depend on $p$. 
Finally, by probing $X_i^t$ with a $p$-layer, that is by checking 
if $X_i^t$ extends down to $p$, we recover $x_i^t$. Although it is easier
to think of one dimensional fragments, this result is valid for any number of
control parameters.

It is interesting to describe a ``traditional'' SOC model with the fragment
language, trying to obtain the $p$-layer description that would make the SOC
model correspond to a usual critical model. 

Let us discuss the one dimensional Bak-Snappen model~\cite{BakSneppen} with 
nearest neighbors interactions. 
In this model one starts from an array of real numbers $a_i$, 
$i=1, \dots, L$ uniformly distributed in the unit interval. One looks for the minimum
of $a_i$ and replaces it and its nearest neighbors with newly generated random numbers, 
again  uniformly distributed in the unit interval. The system auto-organizes so that the 
distribution of $a_i$ follows a power-law (with exponent 1), with a  non-trivial
avalanche distribution. 
We define now a parallel version of the previous model (which is not very efficient
from a computational point of view). For the sake of simplicity, we
 divide the discussion in
two parts: the research of the minimum and the actual evolution. 

Let us visualize the $a_i$ as the lower extremum of segments (fragments) 
$X_i$, and 
 cut the configuration with a line at height $p$. 
The minimum is localized at site $k$ for which  
there is only one intersection. It can be expressed using a Boolean 
variable $\delta_{i,k}^t$ (a Kronecker $\delta$)
\begin{equation}
\delta_{i,k}^t = \bigvee_p x_i^t(p) \left(\bigwedge_{j \neq i} 
    \overline{x_j^t(p)}\right),
\label{min}
\end{equation}
assuming that the minimum is unique 
in the continuous  $p $ limit. 

The fragments $X_i$ at and nearest to the minimum are 
replaced by segments of random length ($R(r)$)
\begin{equation}
X_i^{t+1} = X_i^t \XOR (\Delta_{i-1,k}^t \OR \Delta_{i,k}^t 
\OR \Delta_{i+1,k}^t) (R(r_i^t) \XOR  X_i^t)
\end{equation}
where the fragment $\Delta_{i,k}^t$ is completely filled if $\delta_{i,k}^t=1$
and completely empty if $\Delta_{i,k}^t=0$.
The evolution can be expressed on a $p$-layer as
\begin{equation}
x_i^{t+1} = x_i^t \XOR (\delta_{i-1,k}^t \OR \delta_{i,k}^t \OR
\delta_{i+1,k}^t) ([r_i^t < p] \XOR  x_i^t).
\end{equation}

Similar but more complex expressions can be found for the invasion percolation
process. One can see that  in these ``traditional'' SOC models there 
are long range space
interactions, and also interactions among $p$-layers (see  Eq.\ (\ref{min})). 
We think that the second ingredient is the most important:
if one knows how some quantity like the density varies with $p$ it is not
difficult to imagine a mechanism that automatically reaches the critical point. 
It is still to be proved that this is the
actual mechanism of SOC. On the other hand, Eq.\ (\ref{SOCsite}) shows that 
there exists space and $p$-local mechanisms that can be classified as 
SOC. 

\section{Conclusions}
The fragment method can be considered both as a recipe for numerical
studies of  phase diagrams and as a mapping from criticality to self organized
criticality. 
For what concerns the first topic, the possibility of having a sketch of 
the phase diagram without huge computation resources is useful in determining
the position of the critical line.
Numerical applications of the fragment method will be presented in future work.
From the theoretical point of view, we think that the formalism presented 
in this work allows a clear characterization of the basic properties of 
self organized models, 
suggesting analogies between usual critical phenomena and self
organized ones.

\acknowledgments
We wish to acknowledge fruitful discussions with P. Grassberger, and R.
Bulajich. Partial economic support from {\em CNR} (Italy), {\em CONACYT}
(Mexico), Project DGAPA-UNAM IN103595, Centro Internacional de Ciencias A.C., 
the {\em Programma Vigoni} of the {\em Conferenza permanente dei
rettori delle universit\`{a} italiane} and the workshop {\em Chaos and
Complexity} at ISI-Villa Gualino under the CE contract n. ERBCHBGCT930295 is
also acknowledged.

\appendix
\section*{The Ising Model}
Let us start by considering the one dimensional Ising model. 
Its reduced Hamiltonian can be written as
\begin{equation}
{\cal H}({\bf x})=-\frac{J}{2}\sum_{i=0}^{L-1}\sigma_i\sigma_{i+1}
\end{equation}
where $\sigma_i = 2x_i-1$, and $x_i = 0,1$. 
We choose the heat bath dynamics, \cite{Herrmann,Wang} for which the
probability $\tau ({\bf x\rightarrow y)}$ of going from a configuration
${\bf x}$ to a configuration ${\bf y}$ that can differ from $\bf x$ in a
certain number $\{i_k\}$ of sites is
\begin{equation}
\tau ({\bf x\rightarrow y)=}\frac{\exp (-{\cal H}({\bf y)})}{\sum'
\exp (-{\cal H}({\bf y')}))}
\label{eq:HeatBath}
\end{equation}
where the sum in the denominator extends over all combinations
of the differing
sites $y_{i_k}$. This transition probability does not depend on $\bf x$
and satisfies the detailed balance principle. 

The configuration $\bf y$ is not limited to differ from $\bf x$ only at one
site: the evolution can be applied in parallel changing all even (or odd)
sites. Since the transition
probabilities do not depend on the previous value of 
the cell, the space-time lattice decouples into two noninteracting
sublattices: one with even sites at even times and odd site at odd times, and
the complementary one. By considering only one sublattice, the neighborhood
 of the one dimensional
Ising model is the same of the Domany-Kinzel model. The kinetic Ising
model is just a totalistic cellular automaton (without adsorbing states).

The local 
transition probabilities $\tau(x_{i-1},x_{i+1} \rightarrow y_i)$
can be computed from Eq.\ (\ref{eq:HeatBath}) considering
a difference in just one site. They are
\begin{eqnarray}
	\tau(0,0\rightarrow 1) &=& p/(1+p)\nonumber\\
	\tau(0,1\rightarrow 1) &=& 1/2 \nonumber\\
	\tau(1,0\rightarrow 1) &=& 1/2\nonumber\\
	\tau(1,1\rightarrow 1) &=& 1/(1+p)\nonumber
\end{eqnarray}
with $p=\exp(-2J)$.

It is convenient to introduce here the totalistic functions
$c_k$ that take the value one if the sum of the variables in the
neighborhood is $k$ and zero otherwise. An efficent way of building these
functions is described in \cite{Bagnoli:Boolean}. For the Domany-Kinzel
neighborhood they are
\begin{eqnarray}
	c_0 &=& \overline{x_- \OR x_+}\nonumber\\
	c_1 &=& x_- \XOR x_+ \nonumber\\
	c_2 &=& x_-  x_+.\nonumber
\end{eqnarray}
The evolution equation for the Ising cellular automaton is
\begin{equation}
x' = [r<p/(1+p)] c_0 \OR [r<1/2] c_1 \OR [r<1/(1+p)] c_2.
\end{equation}
 
Before using the fragment method one has to invert the $[r<f(p)]$ ($[r>f(p)]$)
expressions and substitute them with $R(f^{-1}(r))$
($\overline R(f^{-1}(r))$); $p$-independent expressions like $[r<1/2]$ trasform
to $R(0)$ or $\overline R(0)$ according with $r$. 
We leave them in the equations
with the assumption that a true value means $R(0)$ and a false value means
$\overline R(0)$.
We finally find that
\begin{equation}
X' = \overline R\left(\frac{r}{1-r}\right) C_0 + [r<1/2] C_1 +
 R\left(\frac{1-r}{r}\right) C_2.
\end{equation}
where now the $C_k$ are fragments.

For the 2D square Ising model with heat bath dynamics, the Hamiltonian is
\begin{equation}
{\cal H}({\bf x})=-\frac{J}{2}\sum_{i,j=0}^{L-1}
\sigma_{i,j}\sigma_{i+1,j}+\sigma_{i,j}\sigma_{i,j+1}
\end{equation}
with $\sigma_{i,j} = 2x_{i,j} -1$. 
The lattice decouples again in two noninteracting sublattices. 

Repeating the procedure as above, one has again a totalistic cellular
automaton. Using the totalistic functions $C_k$ of four nearest neighbors,
we find that
\begin{eqnarray}
X' &=& R\left(\sqrt{\frac r{1-r}}\right)C _0
	   \OR R\left(\frac r{1-r}\right)C_1
	   \OR \left[r < \frac 12\right] C_2 \nonumber\\
	&& \OR \overline R\left(\frac{1-r}r\right) C_3
	   \OR \overline R\left(\sqrt{\frac{1-r}r}\right) C_4
\label{Ising2D}
\end{eqnarray}
where $p=\exp(-2J)$.

The $C_k$ functions can be computed 
efficiently using the homogeneous polynomials 
$D_j$~\cite{Bagnoli:Boolean}
\begin{eqnarray}
	C_0 &=&\overline{C_1 \OR C_2\OR C_3\OR C_4} \nonumber\\
	C _1 &=&D _1\oplus D _3 \nonumber\\
	C _2 &=&D_2\oplus D _3  \nonumber\\
	C _3 &=&D _3 \nonumber\\
	C _4 &=&D _4 \nonumber
\end{eqnarray}
where
\begin{eqnarray}
 D_1 &=&X _{++}\oplus X_{+-}\oplus X_{-+}\oplus X_{--} \nonumber\\
 D_2 &=&X_{++}X_{+-}\oplus \left(X_{++}\oplus X_{+-}\right)
        \left( X _{-+}\oplus X_{--}\right) \oplus X_{-+}x_{--} \nonumber\\
 D_3 &=&D_2 D_1 \nonumber\\
 D _4 &=&X_{++}X_{+-}X_{-+}X _{--}.\nonumber
\end{eqnarray}

\begin{references}

\bibitem[*]{fb} also INFN and INFM sez. di Firenze; DRECAM-SPEC,
CEA Saclay, 91191 Gif-Sur-Yvette Cedex, France;
e-mail:bagnoli@dma.unifi.it

\bibitem[**]{rr}  On leave from
Facultad de Ciencias, UNAM, Mexico.

\bibitem{Sornette1}  D. Sornette, A. Johansen, and I. Dornic, 
 J. Phys. I (France) {\bf 5}, 325 (1995).

\bibitem{Maslov} S. Maslov and Y.-C. Zhang, Phys. Rev. Let. {\bf 75}, 1550 
 (1995).

\bibitem{PMB} M. Paczuski, S. Maslov, and P. Bak, Phys. Rev. E {\bf 53},
 414 (1996).

\bibitem{Sornette2} D. Sornette and I. Dornic, Phys. Rev. E {\bf 54},
 3334 (1996).

\bibitem{Grassberger} P. Grassberger and Y.-C. Zhang, 
 Physica A {\bf 224}, 169 (1996).

\bibitem{SOC} P. Bak, C. Tang, and K. Wiesenfeld, Phys. Rev. Lett {\bf 59},
 381 (1987).  One can find an extended listing of articles on several aspects
 of SOC in the previuosly cited papers.

\bibitem{Grassbergerorig}  The original presentation occurred at ISI-Villa
 Gualino during the workshop {\it Chaos and Complexity} (Torino, Italy 1994).
 See also refs.~\cite{Roux,Sornette1,Grassberger}.

\bibitem{DP}  J.W.Essam, A.J. Guttmann, and K. de'Bell, J. Phys. A {\bf 16},
 3815 (1983).

\bibitem{Invasion}  D. Wilkinson and J.F. Willemsen, J. Phys. A: Math. Gen. 
 {\bf 16}, 3365 (1983).

\bibitem{BakSneppen}  P. Bak and K. Sneppen, Phys. Rev. Lett. {\bf 71}, 4083
 (1993).

\bibitem{DK}  E. Domany and W. Kinzel, Phys. Rev. Lett. {\bf 53}, 311 (1984).
 W. Kinzel, Z. Phys. B {\bf 58}, 229 (1985).

\bibitem{Georges} A. Georges and P. Le Doussal, J. Stat. Phys. {\bf 54},
 1011 (1989).

\bibitem{multispin} H. J. Herrmann, J. Stat. Phys. {\bf 2},
 145 (1991); D. Stauffer, J. Phys. A: Math. Gen. {\bf 24} L691 (1991).

\bibitem{Iverson} K. Iverson, {\em A Programming Language\/},
	(Wiley, New York, 1962).

\bibitem{Knuth} R. L. Graham, D. E. Knuth, and O.Patashnik,
 {\it Concrete Mathematics}, (Addison Wesley, New York, 1989).

\bibitem{Bagnoli:damage} F. Bagnoli, J. Stat. Phys. {\bf 85}, 151 (1996).

\bibitem{Bagnoli:PRG} F. Bagnoli, R. Bulaijch,  R. Livi, and A. Maritan,
J. Phys. A:Math. Gen. {\bf 25}, L1071 (1992).

\bibitem{ComplexityOfBoolean} I. Wegener, {\em The Complexity of Boolean 
	Functions\/} (Wiley, New York, 1987).

\bibitem{Bagnoli:Boolean} F. Bagnoli, Int. J. of Mod. Phys. C,
 {\bf 3}, 307 (1992).

\bibitem{Roux} S. Roux, A. Hansen, and E. Guyon, J. Phys. (Paris) {\bf 48},
 2125 (1987).

\bibitem{Halpin} T. Halpin-Healy and Y.-C. Zhang, Phys. Rep. {\bf 254}, 
215 (1995). 

\bibitem{linux} {\tt http://www.cs.helsinki.fi/linux}

\bibitem{Herrmann} A. M. Maritz, H. L. Herrmann, and L. de Arcangelis, 
	J. Stat. Phys. {\bf 59}, 1043 (1990).

\bibitem{Wang} F. Wang and M. Suzuki, Physica A {\bf 220}, 534 (1995).
\end{references}
\end{document}